\documentclass[11pt,a4paper]{article}
\pdfoutput=1
\usepackage{jheppub}
\usepackage{rotating}
\usepackage{array}
\usepackage{amsmath}
\usepackage[normalem]{ulem}
\usepackage{slashed}
\usepackage{booktabs}
\usepackage[pdftex,table]{xcolor}
\usepackage{units}
\usepackage{mathrsfs}
\usepackage{dsfont}
\usepackage{amsmath}
\usepackage{txfonts}
\usepackage{amssymb}
\renewcommand{\vec}[1]{\mathbf{#1}}

\usepackage{multirow}

\preprint{DESY 18-015, TUM-HEP 1130/18,  KIAS-P18008}
\title{Halo-independent comparison of direct detection experiments in the effective theory of dark matter-nucleon interactions}

\author[a]{Riccardo Catena}
\author[b,c]{Alejandro Ibarra}
\author[b]{Andreas Rappelt}
\author[d]{Sebastian Wild}
\affiliation[a]{Chalmers University of Technology, Department of Physics, SE-412 96 G\"oteborg, Sweden}
\affiliation[b]{Physik-Department T30d, Technische Universit\"at M\"unchen,\\James-Franck-Stra\ss{}e, 85748 Garching, Germany}
\affiliation[c]{School of Physics, Korea Institute for Advanced Study, Seoul 02455, South Korea}
\affiliation[d]{DESY, Notkestra\ss e 85, D-22607 Hamburg, Germany}
\emailAdd{catena@chalmers.se}
\emailAdd{ibarra@tum.de}
\emailAdd{andreas.rappelt@tum.de}
\emailAdd{sebastian.wild@desy.de}

\abstract{The theoretical interpretation of dark matter direct detection experiments is hindered by uncertainties of the microphysics governing the dark matter-nucleon interaction, and of the dark matter density and velocity distribution inside the Solar System. These uncertainties are especially relevant when confronting a detection claim to the null results from other experiments, since seemingly conflicting experimental results may be reconciled when relaxing the assumptions about the form of the interaction and/or the velocity distribution. We present in this paper a  halo-independent method to calculate the maximum number of events in a direct detection experiment given a set of null search results, allowing for the first time the scattering to be mediated by an arbitrary combination of various interactions (concretely we consider up to 64). We illustrate this method to examine the compatibility of the dark matter interpretation of the three events detected by the silicon detectors in the CDMS-II experiment with the null results from XENON1T and PICO-60.
  }

\begin{document}
\maketitle

\section{Introduction}
\label{sec:introduction}

Various cosmological and astronomical observations provide evidence for the existence of a non-luminous matter component in our Universe, commonly dubbed dark matter (DM). The whole body of observations can be qualitatively well reproduced if the dark matter is constituted by new particles not contained in the Standard Model (see \emph{e.g.}~\cite{Bertone:2004pz} for a review). This hypothesis could be tested if dark matter particles have sizable interactions with nucleons~\cite{Goodman:1984dc}. If this is the case, dark matter particles traversing a detector located at the Earth could scatter off the nuclei in the target material, producing in their recoil a potentially detectable signal in the form of ionization, scintillation light or temperature rise. However, similar signals can also be generated by electromagnetic interactions of $\alpha$-particles, electrons, and photons produced by the radioactive isotopes in the surrounding material, as well as by nuclear interactions of neutrons produced by natural radioactivity. Therefore, attributing the observed nuclear recoils to an interaction with a dark matter particle requires an excellent understanding of the experimental set-up and, in any case, requires independent confirmation from other experiments, preferably using different target materials.

Several experiments are searching for such rare events. XENON1T~\cite{Aprile:2017iyp}, PandaX~\cite{Cui:2017nnn} and LUX~\cite{Akerib:2016vxi} use xenon as target material, PICO-60~\cite{Amole:2017dex} employs octafluoropropane (${\rm C}_3 {\rm F}_8$), CDMS-II uses germanium~\cite{Ahmed:2010wy} and silicon~\cite{Agnese:2013rvf}, while CRESST-II~\cite{Angloher:2015ewa} is based on a ${\rm Ca W O}_4$ target. Experiments using xenon, ${\rm C}_3 {\rm F}_8$, germanium or ${\rm Ca W O}_4$ as target materials have found no evidence for dark matter. On the other hand, the search for dark matter particles with the silicon detectors of the CDMS-II experiment revealed three dark matter candidate events, which can be attributed to a statistical fluctuation of the (known) backgrounds, to an unaccounted for background, or to a dark matter signal. If interpreted as a dark matter signal, the parameters necessary to explain the three observed events are in strong tension with the results from null search experiments, assuming elastic scattering of dark matter with the nucleus mediated by the spin-independent (SI) or spin-dependent (SD) interaction only, and assuming a Maxwell-Boltzmann velocity distribution for the dark matter particles. 

The origin of the three events remains a mystery  to this day, as no new backgrounds have been identified which could explain the observed nuclear recoils, and the probability that the signal is due to a statistical fluctuation is smaller than 5.4\%~\cite{Agnese:2013rvf}. It is then worthwhile investigating whether the CDMS-Si signal can be reconciled with the various null search experiments by considering other frameworks of dark matter scattering with the nuclei and/or by relaxing the assumption that the dark matter velocity distribution inside the Solar System has the Maxwell-Boltzmann form.

Motivated by the CDMS-Si results, we propose in this paper a method to confront, in a halo-independent manner, a putative dark matter signal with the null results from other direct detection experiments in an effective theory of dark matter-nucleon interactions, including possible interferences among them. The method then allows to establish the compatibility among different experiments, regardless of our current ignorance of the concrete form of the local dark matter velocity distribution, and (practically) regardless of the particle physics characteristics of the dark matter-nucleon interaction.  This extends previous works along this direction which have either conducted a halo-independent analysis for fixed assumptions regarding the particle physics nature of dark matter (being it the standard SI or SD interaction or any non-standard form of dark matter-nucleon interaction)~\cite{Frandsen:2013cna,DelNobile:2013cva,Fox:2014kua,Scopel:2014kba,Feldstein:2014ufa,Bozorgnia:2014gsa,Anderson:2015xaa,Gelmini:2016pei,Gondolo:2017jro,Ibarra:2017mzt}, or an analysis in the full parameter space of the non-relativistic effective theory of dark matter-nucleon interactions for a fixed choice of the dark matter velocity distribution~\cite{Catena:2016hoj}.  We illustrate our method confronting the tentative detection of a dark matter signal by the silicon detectors of the CDMS-II experiment with the null results from the XENON1T and PICO-60 experiments, although the method has general applicability.

The paper is organized as follows. In Section~\ref{sec:theory} we review the non-relativistic effective theory of dark matter-nucleon interactions for a general velocity distribution. In Section~\ref{sec:method} we present our method to confront a detection claim with null search results in a halo independent manner and for a large class of operators inducing dark matter-nucleon interactions, and apply our method to assess the viability of the dark matter interpretation of the CDMS-II signal in the silicon detectors. The resulting upper limits on the expected number of dark matter induced events are discussed in Section~\ref{sec:results}. Finally, in Section~\ref{sec:conclusions} we present our conclusions. We also include two appendices with an analytical derivation of the velocity distribution that maximizes the event rate at the CDMS-Si detector, and of the implementation of the direct detection experiments employed in our analysis.

\section{Non-relativistic effective theory for arbitrary velocity distributions}
\label{sec:theory}
In the non-relativistic scattering of DM particles off nuclei in terrestrial detectors, the ratio of momentum transfer to the constituent nucleon mass is smaller than one, i.e. $|\vec q|/m_{N} \ll 1$. Consequently, the amplitude for DM scattering off nucleons $N$ in target nuclei, $\mathscr{M}_{\chi N}$, can in general be expanded in powers of $|\vec q|/m_{N}$. Based on general symmetry arguments, each term in this expansion must be invariant under Galilean transformations and Hermitian conjugation, and can be expressed in terms of basic invariants under the above symmetries~\cite{Fan:2010gt,Fitzpatrick:2012ix}:  $i\vec q$, $\vec v^\perp \equiv \vec u + \vec q/2 \mu_N$, $\vec S_\chi$, and $\vec S_N$, where $\vec q$ is the momentum transfer, $\mu_N$ and $\vec u$ are the DM-nucleon reduced mass and relative velocity, respectively, and $\vec S_{\chi}$ ($\vec S_N$) is the DM (nucleon) spin. From $\mathscr{M}_{\chi N}$, one can construct the associated Hamiltonian for the interaction of DM with a nucleus $T$, $\mathscr{H}_{\chi T}$, which in the one-body approximation is given by~\cite{Fan:2010gt,Fitzpatrick:2012ix}
\begin{equation}
\mathscr{H}_{\chi T} = \sum_{i}\sum_{j} \Big( c_j^0  \hat{\mathcal{O}}^i_j \, \mathds{1}_{2\times 2}^i + c_j^1 \, \hat{\mathcal{O}}^i_j \, \mathds{\tau}_{3}^i \Big)
\label{eq:H}
\end{equation}
where the index $j$ characterises the DM-nucleon interaction type, $c_j^0$ ($c_j^1$) is the associated isoscalar (isovector) coupling constant, the $A$ nucleons in the target nucleus are labeled by the index $i=1,\dots,A$, $\mathds{1}_{2\times 2}^i$ ($\mathds{\tau}_{3}^i$) is the identity (third Pauli matrix) in the $i$-th nucleon isospin space, and $\hat{\mathcal{O}}^i_j$ is a non-relativistic operator for interactions of type $j$ between DM and the $i$-th nucleon. The operators $\hat{\mathcal{O}}^i_j$ act on particle coordinates, once momentum transfer and transverse relative velocity are quantised and expressed in terms of the operators $\hat{\vec q}$ and $\hat{\vec v}^\perp$, respectively. After quantisation, the DM and nucleon spin operators are denoted by $\hat{\vec S}_\chi$ and $\hat{\vec S}_N$, respectively. At linear order in $\hat{\vec v}^{\perp}$, there are 16 independent interaction types $\hat{\mathcal{O}}^i_j$, although not all of them appear as leading operators in the non-relativistic limit of simplified models. The 16 operators are listed in Tab.~\ref{tab:operators}. $\hat{\mathcal{O}}_{17}$ and $\hat{\mathcal{O}}_{18}$ only arise for spin 1 DM~\cite{Dent:2015zpa}. In the notation of~\cite{Fitzpatrick:2012ix}, $\hat{\mathcal{O}}_{2}$ is quadratic in $\vec v^\perp$, and $\hat{\mathcal{O}}_{16}$ is a linear combination of $\hat{\mathcal{O}}_{12}$ and $\hat{\mathcal{O}}_{15}$.~For these reasons $\hat{\mathcal{O}}_{2}$ and $\hat{\mathcal{O}}_{16}$ are not shown in Tab.~\ref{tab:operators}, and will not be considered here.
\setlength{\tabcolsep}{24pt}
\begin{table}[t]
    \centering
    \begin{tabular}{ll}
	 \toprule
	  \toprule
        $\hat{\mathcal{O}}_1 = \mathds{1}_{\chi}\mathds{1}_N$  & $\hat{\mathcal{O}}_{10} = i{\bf{\hat{S}}}_N\cdot\frac{{\bf{\hat{q}}}}{m_N}\mathds{1}_\chi$\\  
        $\hat{\mathcal{O}}_3 = i{\bf{\hat{S}}}_N\cdot\left(\frac{{\bf{\hat{q}}}}{m_N}\times{\bf{\hat{v}}}^{\perp}\right)\mathds{1}_\chi$ & $\hat{\mathcal{O}}_{11} = i{\bf{\hat{S}}}_\chi\cdot\frac{{\bf{\hat{q}}}}{m_N}\mathds{1}_N$ \\
        $\hat{\mathcal{O}}_4 = {\bf{\hat{S}}}_{\chi}\cdot {\bf{\hat{S}}}_{N}$ & $\hat{\mathcal{O}}_{12} = {\bf{\hat{S}}}_{\chi}\cdot \left({\bf{\hat{S}}}_{N} \times{\bf{\hat{v}}}^{\perp} \right)$  \\                                                      
        $\hat{\mathcal{O}}_5 = i{\bf{\hat{S}}}_\chi\cdot\left(\frac{{\bf{\hat{q}}}}{m_N}\times{\bf{\hat{v}}}^{\perp}\right)\mathds{1}_N$ & $\hat{\mathcal{O}}_{13} =i \left({\bf{\hat{S}}}_{\chi}\cdot {\bf{\hat{v}}}^{\perp}\right)\left({\bf{\hat{S}}}_{N}\cdot \frac{{\bf{\hat{q}}}}{m_N}\right)$\\       
        $\hat{\mathcal{O}}_6 = \left({\bf{\hat{S}}}_\chi\cdot\frac{{\bf{\hat{q}}}}{m_N}\right) \left({\bf{\hat{S}}}_N\cdot\frac{\hat{{\bf{q}}}}{m_N}\right)$& $\hat{\mathcal{O}}_{14} = i\left({\bf{\hat{S}}}_{\chi}\cdot \frac{{\bf{\hat{q}}}}{m_N}\right)\left({\bf{\hat{S}}}_{N}\cdot {\bf{\hat{v}}}^{\perp}\right)$\\  
        $\hat{\mathcal{O}}_7 = {\bf{\hat{S}}}_{N}\cdot {\bf{\hat{v}}}^{\perp}\mathds{1}_\chi$ & $\hat{\mathcal{O}}_{15} = -\left({\bf{\hat{S}}}_{\chi}\cdot \frac{{\bf{\hat{q}}}}{m_N}\right)\left[ \left({\bf{\hat{S}}}_{N}\times {\bf{\hat{v}}}^{\perp} \right) \cdot \frac{{\bf{\hat{q}}}}{m_N}\right] $\\ 
        $\hat{\mathcal{O}}_8 = {\bf{\hat{S}}}_{\chi}\cdot {\bf{\hat{v}}}^{\perp}\mathds{1}_N$  & $\hat{\mathcal{O}}_{17}=i \frac{{\bf{\hat{q}}}}{m_N} \cdot \mathbf{\mathcal{S}} \cdot {\bf{\hat{v}}}^{\perp} \mathds{1}_N$\\ 
        $\hat{\mathcal{O}}_9 = i{\bf{\hat{S}}}_\chi\cdot\left({\bf{\hat{S}}}_N\times\frac{{\bf{\hat{q}}}}{m_N}\right)$ & $\hat{\mathcal{O}}_{18}=i \frac{{\bf{\hat{q}}}}{m_N} \cdot \mathbf{\mathcal{S}}  \cdot {\bf{\hat{S}}}_{N}$ \\
    \bottomrule
      \bottomrule
    \end{tabular}
    \caption{Operators defining the non-relativistic effective theory of DM-nucleon interactions~\cite{Fan:2010gt,Fitzpatrick:2012ix}.~The notation is the one introduced in Sec.~\ref{sec:theory}.~The matrixes $\mathds{1}_N$ and $\mathds{1}_\chi$ represent the identity in the nucleon and DM spin space, respectively. The operators $\hat{\mathcal{O}}_{1}$ and $\hat{\mathcal{O}}_{4}$ correspond to canonical spin-independent and spin-dependent interactions, respectively.~The operators $\hat{\mathcal{O}}_{17}$ and $\hat{\mathcal{O}}_{18}$ only arise for spin 1 WIMPs, with $\mathbf{\mathcal{S}}$ being a symmetric combination of spin 1 DM polarisation vectors~\cite{Dent:2015zpa}.~Operator $\hat{\mathcal{O}}_{2}$ is quadratic in ${\bf{\hat{v}}}^{\perp}$ and $\hat{\mathcal{O}}_{16}$ is a linear combination of $\hat{\mathcal{O}}_{12}$ and $\hat{\mathcal{O}}_{15}$ and are therefore not considered here~\cite{Anand:2013yka}.}
    \label{tab:operators}
\end{table}

The differential cross-section for DM-nucleus scattering, ${\rm d}\sigma_T/{\rm d} E_R$, quadratically depends on nuclear matrix elements of $\mathscr{H}_{\chi T}$ and on the coupling constants $c_j^\tau$ in Eq.~(\ref{eq:H}). For an explicit expression, see, e.g.~\cite{Anand:2013yka}. Unlike~\cite{Anand:2013yka}, however, here we also include contributions to ${\rm d}\sigma_T/{\rm d} E_R$ from long-range interactions by replacing the coupling constants $c_j^\tau$ in Eq.~(\ref{eq:H}) with the coefficients $c_j^\tau + \tilde{c}_j^\tau \cdot m_N^2/|\vec q|^2$, where $c_j^\tau$ and $\tilde c_j^\tau$ can vary independently. In total, our effective theory contains 64 Wilson coefficients, corresponding to the 16 operators describing the scattering  off nucleons, each one with an isoscalar and isovector component, and each one in the form of contact and long-range interactions. This very general parameterization of the DM scattering process then captures almost all conceivable particle physics scenarios for the interaction of DM with nucleons. Important exceptions are models where DM scatters inelastically off nuclei, or scenarios involving a mediator with a mass $m_\text{med} \simeq |\vec q| \simeq \,(1-100)\,$MeV, which neither corresponds to the short- nor to the long-range version of the operators considered in this work.

Nuclei of interest for the present analysis are carbon, fluorine, silicon and xenon. We calculate ${\rm d}\sigma_T/{\rm d} E_R$ using standard nuclear physics methods reviewed in~\cite{Fitzpatrick:2012ix,Catena:2015uha}. This calculation requires one-body density matrix elements (OBDMEs) for all atomic nuclei listed above. Here we use OBDMEs computed in~\cite{Anand:2013yka} and implemented in the {\sffamily Mathematica} package {\sffamily DMFormFactor}. In order to validate our findings, we also perform an independent shell-model calculation of the OBDMEs for $^{29}$Si, using the so-called w-interaction in the sd valence space as an input~\cite{Warburton:1992rh}. We find perfect agreement between our results and the ones reported in~\cite{Anand:2013yka}.

Finally, one can calculate the differential rate of signal events per unit detector mass in a given experiment as follows:
\begin{align}
\frac{{\rm d} R}{{\rm d} E_R} = \sum_T \xi_T \frac{\rho_\chi}{m_\chi m_T} \int_{|\vec v| \ge v_{\rm min}} {\rm d}^3 v \, |\vec v| f(\vec v) \, \frac{{\rm d}\sigma_T}{{\rm d} E_R} (v^2, E_R)
\label{eq:rate}
\end{align}
where $m_\chi$ is the DM particle mass, $\rho_\chi$ is the local DM density, $v_{\rm min}=\sqrt{2 m_T E_R}/(2 \mu_T)$ is the minimum DM velocity required to deposit an energy $E_R$ in the detector, $\mu_T$ and $m_T$ are the DM-nucleus reduced mass and target nucleus mass, respectively, and the sum is extended to all elements in the detector material, each one with a mass fractions $\xi_T$. In the case of the XENON1T experiment, we consider the seven most abundant Xenon isotopes, whereas for the CDMS-Si experiment we include two isotopes, namely $^{28}$Si and $^{29}$Si. Furthermore, $f$ denotes the DM velocity distribution in the \emph{detector} rest frame, which is related to the galactic velocity distribution $f_G$ via $f(\vec v) = f_G(\vec v + \vec v_\oplus(t))$, with $\vec v_\oplus(t)$ being the Earth velocity in the galactic rest frame. As we only consider experiments measuring the total rate of nuclear recoils (as opposed to the annual modulation of the signal rate), we will neglect the resulting small time dependence of the velocity distribution $f$ in the detector rest frame. Integrating Eq.~(\ref{eq:rate}) above the appropriate threshold energy (accounting for finite efficiency and energy resolution) we find the expected number of signal events in a given experiment. Further details on the implementation of XENON1T, PICO-60 and CDMS-Si can be found in Appendix~\ref{sec:exp}. 

For a given velocity distribution $f(\vec v)$, the number of expected signal events in an experiment ${\cal E}$ can be written as
\begin{align}
N_{f(\vec v)}^{({\cal E})} \left( {\bf c }\right) = {\bf c}^T \mathbb{N}_{f(\vec v)}^{({\cal E})} {\bf c} \,.
\label{eq:Nf_quadratic}
\end{align}
Here, ${\bf c}$ is the vector of all coefficients $c_j^\tau$ and $\tilde{c}_j^\tau$, and $\mathbb{N}_{f(\vec v)}^{({\cal E})}$ is a real symmetric matrix. The matrix $\mathbb{N}_{f(\vec v)}^{({\cal E})}$ encodes all the information about the nuclear response functions upon which ${\rm d}\sigma_T/{\rm d} E_R$ depends, the velocity distribution $f$, and the details of the experiment ${\cal E}$. It is however independent of the coefficient vector $\mathbf{c}$. 

\section{Confronting the CDMS-Si signal to null search results}
\label{sec:method}

The silicon detectors of the CDMS-II experiment have revealed three dark matter candidate events with an exposure of $140.2$ kg day~\cite{Agnese:2013rvf}. On the other hand, neither the XENON1T experiment, with an exposure of $3.35\times 10^4$ kg day~\cite{Aprile:2017iyp}, nor the PICO-60 experiment, with an exposure of $1167$ kg day~\cite{Amole:2017dex} have found evidence for dark matter, despite their larger exposures\footnote{In the following, we will only discuss the implications of those two null result experiments on the potential dark matter signal in CDMS-Si. In particular, we have checked that employing the CRESST-II~\cite{Angloher:2015ewa} results instead of the one by the PICO collaboration leads to less stringent bounds on the signal strength in CDMS-Si.}. Assuming elastic DM-nucleus scattering mediated by the spin-independent interaction only (or by the spin-dependent interaction only), and assuming a Maxwell-Boltzmann velocity distribution for the dark matter particles, the CDMS-Si results are largely incompatible with the negative searches by XENON1T and PICO-60. On the other hand, it is worthwhile noting that these three experiments employ different target materials, with responses which can be dramatically affected by the specific nature of the DM-nucleon interaction. Furthermore, the typical energy of the nuclear recoils, and whether they are detectable or not, depends crucially on the dark matter velocity distribution. It then remains an open question whether the CDMS-Si signal can be reconciled with the various null search experiments by considering other frameworks of dark matter scattering with the nuclei and/or by relaxing the assumption that the dark matter velocity distribution has the Maxwell-Boltzmann form. 

Rather than sampling the large (virtually infinite) number of possibilities offered by model building, here we pursue a halo-independent approach to the effective theory of dark matter-nucleus interactions, which encompasses all halo models and a large number of possible interactions, possibly interfering with each other. Concretely, in our approach we  calculate, for a fixed dark matter mass, the maximal number of DM induced recoil events at CDMS-Si for all possible (normalized to unity)  velocity distributions and for all possible values of the Wilson coefficients $\bf c$. Namely: 
\begin{align}
N_\text{max}^{(\text{CDMS-Si})} &\equiv  \max_{f(\vec v)} \max_{\bf c} \left[ N_{f(\vec v)}^{(\text{CDMS-Si})} \left( {\bf c }\right) \right]\,, \\
{\rm subject~to}&~~~~~  N_{f(\vec v)}^{(\text{XENON1T})}(\vec c)\leq N^{\rm (XENON1T)}_{\rm u.l.}\,, \\
{\rm and }&~~~~~  N_{f(\vec v)}^{(\text{PICO})}(\vec c)\leq N^{\rm (PICO)}_{\rm u.l.}\,, \\
{\rm and }&~~~~~\int f(\vec v)\,{\rm d}^3 v =1\,,
\label{eq:Nmax_definition}
\end{align}
where $N^{\rm (XENON1T)}_{\rm u.l.}$ and $N^{\rm (PICO)}_{\rm u.l.}$ are, respectively, the 95\% C.L. upper limits on the number of events at XENON1T and PICO-60. The dark matter interpretation of the three CDMS-Si events will be disfavored for that specific dark matter mass if  $N_\text{max}^{(\text{CDMS-Si})} \ll 3$. To this end, we determine $N_\text{max}^{(\text{CDMS-Si})}$ in two steps. We first determine the maximum number of events at CDMS-Si for a fixed velocity distribution, and then we select the maximum possible number of events from sampling over all possible velocity distributions.

\begin{figure}
	\begin{center}
		\includegraphics[scale=1.0]{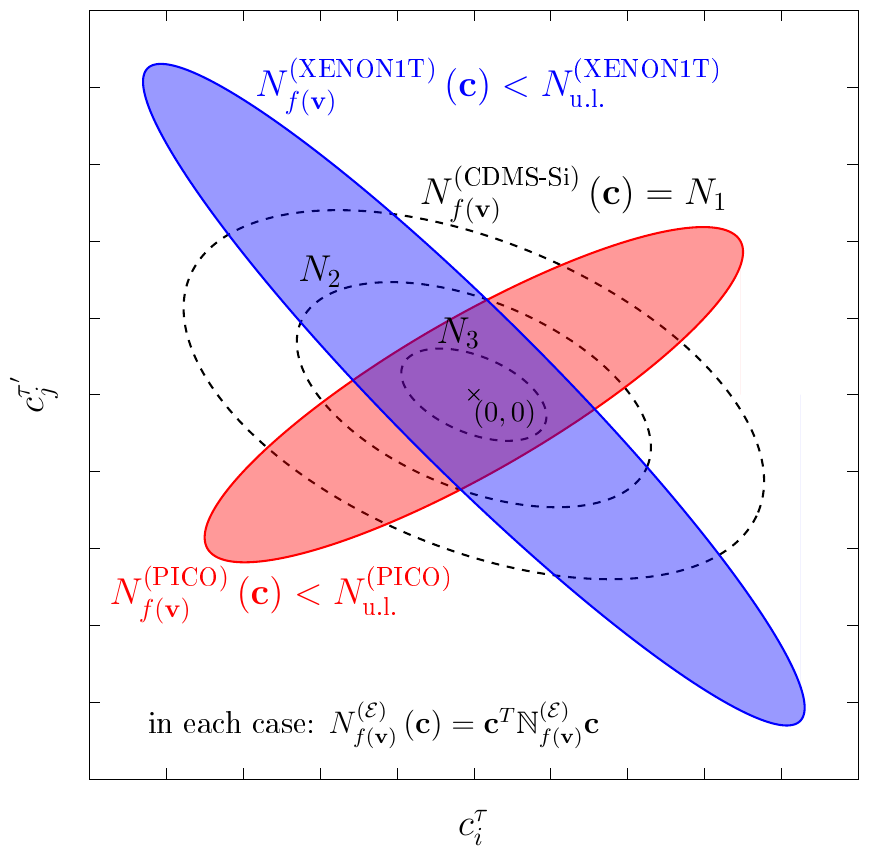}
	\end{center}
	\caption{Schematic description of our method to determine the maximal number of events expected at CDMS-Si which is compatible with the constraints from XENON1T and PICO-60 for a given velocity distribution $f(\vec v)$. In this example (which is not based on real data), we consider two Wilson coefficients $c_i^\tau$ and $c_j^{\tau'}$, and show for each of the three experiments the ellipses corresponding to a fixed number of expected events. In this case, the maximally allowed number of events at CDMS-Si would be equal to $N_2$; see the text for details.}
	\label{fig:ellipses_example}
\end{figure}

In the first step of our approach we fix the dark matter velocity distribution and we maximize, for a given dark matter mass, the number of events at CDMS-Si with the constraints from XENON1T and PICO-60. In our analysis we include up to 64 different interactions, which makes the determination of ${\rm max}_{\bf c} N_{f(\vec v)}^{(\text{CDMS-Si})}(\bf c)$ using scanning techniques unfeasible. Instead, we pursue the semi-analytic approach proposed in~\cite{Catena:2016hoj}, which is based  on the geometric interpretation of Eq.~(\ref{eq:Nf_quadratic}).  For an experiment ${\cal E}$ reporting a null result, the allowed values of the parameters ${\bf c}$ are defined by the condition  $N_{f(\vec v)}^{({\cal E})} \left( {\bf c }\right) \leq N_{\rm u.l.}^{({\cal E})}$, which geometrically corresponds to the interior of an ellipsoid in $\text{dim}({\bf c})$ dimensions.
This is illustrated in Fig.~\ref{fig:ellipses_example} for the slice of the ${\bf c}$-parameter space spanned by the two Wilson coefficients $c_i^\tau$ and $c_j^{\tau'}$. All parameter points ${\bf c}$ within the blue ellipse are allowed by the (in this case hypothetical) data from the XENON1T experiment, while the red ellipse corresponds to the bound set by PICO-60; the region of the parameter space allowed by both experiments is then determined by the intersection of both regions. The maximum number of events at CDMS-Si compatible with the XENON1T and PICO-60 results is determined by the largest ellipse ${\bf c}^T \mathbb{N}_{f(\vec v)}^{\rm (CDMS-Si)} {\bf c}=N_{f(\vec v)}^{\rm (CDMS-Si)}$ which has a non-empty intersection with the region of the parameter space allowed by the two null results. This is also illustrated in the Figure, which shows, as dashed black lines, isocontours of $N^{\rm (CDMS-Si)}_{f(\vec v)}$ with increasing larger number of events, {\it i.e.} $N_1>N_2>N_3$. A number of events at CDMS-Si equal to $N_1$ is in conflict with the null search experiments, equal to $N_3$ is allowed, and equal to $N_2$ is marginally allowed. Therefore, for this concrete velocity distribution, the maximal number of events compatible with the null search experiments is equal to $N_2$.
\pagebreak

More specifically, to determine whether the ellipse ${\bf c}^T \mathbb{N}_{f(\vec v)}^{\rm (CDMS-Si)} {\bf c}=N_{f(\vec v)}^{\rm (CDMS-Si)}$ intersects the region of the parameter space allowed by XENON1T and PICO-60, it is sufficient to find non-negative real parameters $\zeta^{\rm (XENON1T)}$ and $\zeta^{\rm (PICO)}$, such that~\cite{Catena:2016hoj}
\begin{align}
&{\it i})\, \,  \zeta^{\rm (XENON1T)} +\zeta^{\rm (PICO)}< 1 \quad \text{and} \nonumber \\
&{\it ii}) \, \, \left(  \zeta^{\rm(XENON1T)} \frac{\mathbb{N}_{f(\vec v)}^{\rm(XENON1T)}}{N_{\rm u.l.}^{\rm (XENON1T)}} +\zeta^{\rm (PICO)} \frac{\mathbb{N}_{f(\vec v)}^{\rm(PICO)}}{N_{\rm u.l.}^{\rm (PICO)}} \right) - \frac{\mathbb{N}_{f(\vec v)}^{\rm (CDMS-Si)}}{N^{\rm (CDMS-Si)}_{f(\vec v)}} \,\, \text{is a positive definite matrix,}
\label{eq:zeta_condition}
\end{align}
where the matrices $\mathbb{N}_{f(\vec v)}^{({\cal E})}$ were defined in Eq.~(\ref{eq:Nf_quadratic}). We then determine ${\rm max}_{\bf c} N^{\rm (CDMS-Si)}_{f(\vec v)}(\bf c)$ as the largest value of $ N^{\rm (CDMS-Si)}_{f(\vec v)}$ for which such a solution exists. \footnote{For the numerical implementation of the algorithm, we use the {\texttt feasp} solver implemented in MATLAB~\cite{MATLAB:2014a}.}

The above construction allows to find the maximal number of events at CDMS-Si compatible with the null searches from XENON1T and PICO-60 for a fixed velocity distribution. The second step in our maximization approach requires to repeat the above procedure for all possible velocity distributions and to determine the absolute maximum. Again, sampling over an (infinitely large) set of functions is unfeasible. On the other hand, it is possible to show (see Appendix \ref{app:vel_distr}) that the velocity distribution that maximizes the number of events at CDMS-Si consists of (at most) two dark matter streams with velocities $\vec v_1$ and  $\vec v_2$, with weights equal to $\alpha$ and $(1-\alpha)$, where $0\leq \alpha\leq 1$. As mentioned previously, in the calculation of the time-integrated total rates we neglect the time dependence of the velocity distribution $f(\vec v)$ in the detector rest frame, such that (for non-directional detectors) it is sufficient to consider the equivalent one-dimensional velocity distribution
\begin{align}
f_{\alpha, v_1, v_2}(v) = \alpha \, \delta(v-v_1) + (1-\alpha) \delta(v-v_2) \,,
\label{eq:fv_form}
\end{align}
which only depends on the three parameters $\alpha$, $v_1=|\mathbf{v}_1|$ and $v_2=|\mathbf{v}_2|$\footnote{For confronting experiments where the time dependence of the dark matter flux at the Earth plays a critical role, as is the case of the annual modulation signal measured by DAMA, such a simplification is not possible.}. Therefore, the maximum number of events can be calculated from
\begin{align}
N_\text{max}^{(\text{CDMS-Si})} &\equiv  \max_{\alpha,  v_1, v_2}\max_{\bf c} \left[ N_{\alpha,  v_1 v_2}^{(\text{CDMS-Si})} \left( {\bf c }\right) \right]\,.
\end{align}
Namely,  we calculate the maximal number of events at CDMS-Si for a sample of velocity distributions of the form Eq.~(\ref{eq:fv_form}), with different values of $\alpha$, $v_1$ and $v_2$,\footnote{Concretely, we take 50 values of $\alpha$, and 100 values of the velocities $v_1$ and $v_2$.} and we determine  $N_\text{max}^{(\text{CDMS-Si})}$ by taking the largest value in the sample.

\begin{figure}
\begin{center}
\hspace*{-0.5cm}
\includegraphics[scale=0.9]{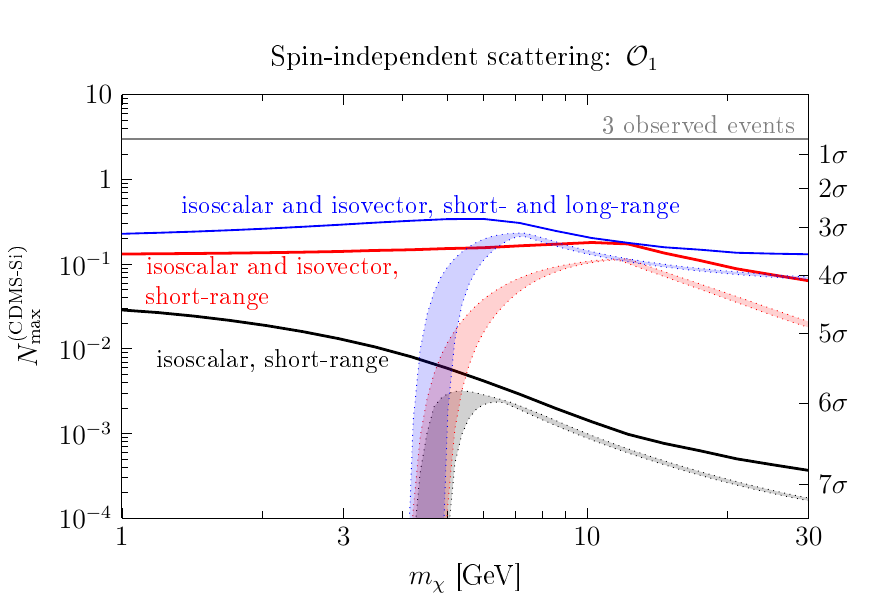}\hspace*{0.1cm}
\includegraphics[scale=0.9]{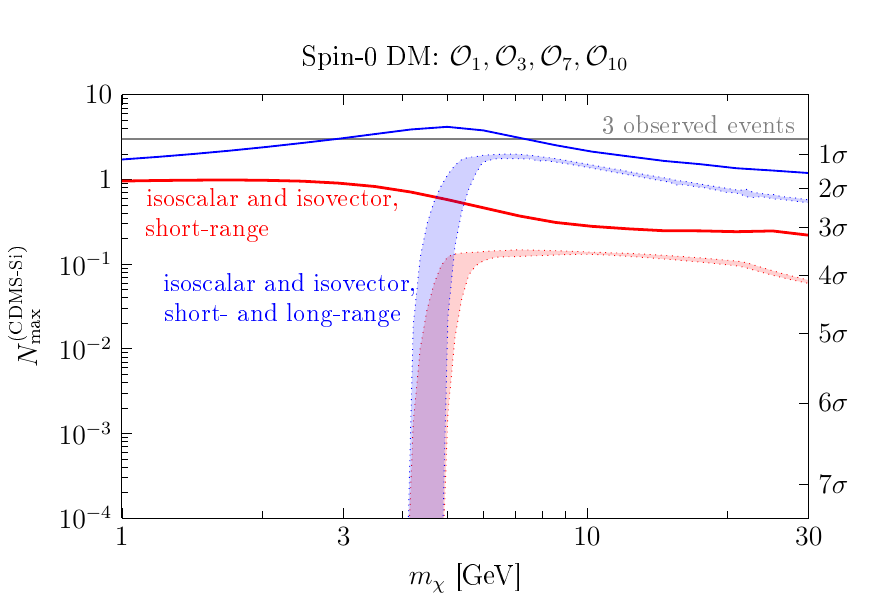}\\
\hspace*{-0.5cm}
\includegraphics[scale=0.9]{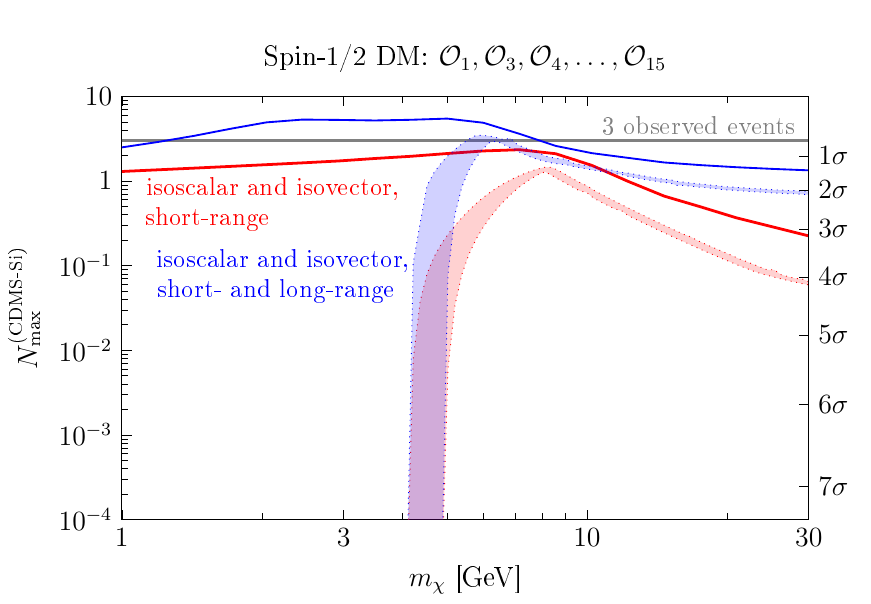}\hspace*{0.1cm}
\includegraphics[scale=0.9]{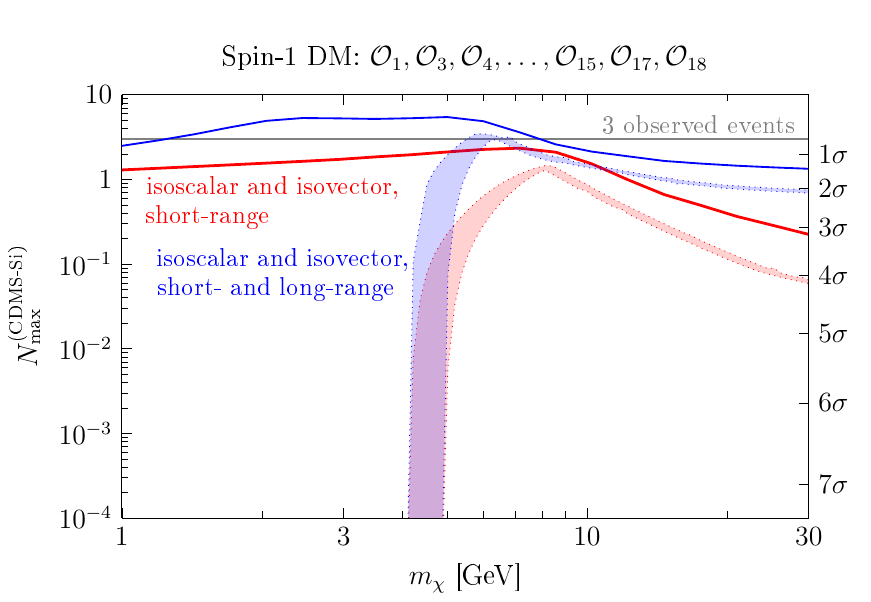}
	\end{center}
    \caption{Maximal number of events at CDMS-Si compatible with the null search results from XENON1T and PICO-60 as a function of the dark matter mass. The four panels correspond to different sets of operators $\hat{\mathcal{O}}_j$ employed in the analysis, while the different colors in each panel show the results depending on whether the isoscalar and/or isovector as well as the short- and/or long-range versions of the operators are included. The solid curves correspond to the halo-independent upper limit on $N_\text{max}^{(\text{CDMS-Si})}$, while the shaded bands show the upper limit assuming a Maxwell-Boltzmann distribution with $v_0$ ranging from $220$ to $240$ km/s and $v_\text{esc}$ from $492$ to $587$ km/s.}  
    \label{fig:main_result}
  \end{figure}

\section{Results}
\label{sec:results}

The maximal number of events at CDMS-Si compatible with the null search results from XENON1T and PICO-60, $N_\text{max}^{(\text{CDMS-Si})}$, is shown in Fig.~\ref{fig:main_result} as a function of the dark matter mass. The upper-left panel refers to the case of scattering mediated by the spin-independent interaction only, while the upper-right, bottom-left and bottom-right panels correspond to scenarios where the dark matter particle has spin 0, spin 1/2 and spin 1, respectively. The shaded bands were obtained assuming a Maxwell-Boltzmann velocity distribution with most probable speed $v_0$ ranging between 220 and 240 km/s and escape velocity $v_{\rm esc}$ between 492 and 587 km/s, while the solid lines were obtained with our halo independent approach. The red color shows results obtained assuming  isoscalar and isovector short-range interactions, while the blue color also includes the possibility of long-range interactions; the black color in the upper-left panel shows results obtained for the commonly studied case of an isoscalar, short-ranged, spin-independent interaction, and is shown as reference. In each panel of Fig.~\ref{fig:main_result}, the right axis shows the significance of exclusion of the dark matter interpretation of the CDMS-Si events. This quantity is defined as the number of standard deviations corresponding to the probability of observing three or more events when expecting only $N_\text{max}^{(\text{CDMS-Si})}$ events, setting for definiteness the number of observed background events to zero. Obviously, the significance of exclusion for a given value of $N_\text{max}^{(\text{CDMS-Si})}$ would decrease if one or more of the observed events were due to the known (or unknown) backgrounds.

In all panels of Fig.~\ref{fig:main_result}, the maximally allowed number of events in CDMS-Si obtained under the assumption of the Maxwell-Boltzmann distribution is zero for $m_\chi \lesssim 4\,$GeV. This is simply a consequence of the fact that the velocity distribution has a cutoff at the galactic escape velocity $v_\text{esc}$, and thus all recoils fall below the CDMS-Si threshold once dark matter is sufficiently light. On the other hand, in the halo-independent approach we do not impose a cutoff on the velocity distribution and thus in principle also lighter dark matter candidates can account for the events observed by CDMS-Si. However, at some point the required velocities get so large that even for highly non-standard galactic dynamics governing the dark matter distribution the corresponding $f(v)$ becomes unphysical.  In light of this, we only show our results for $m_\chi > 1\,$GeV, corresponding to a minimal velocity $v_\text{min} \simeq 3000\,$km/s required for observing recoils in CDMS-Si\footnote{Let us also remark that, at least for velocity-independent interactions, in the limit $m_\chi \ll m_T$ the recoil rate corresponding to a stream with velocity $v_0$ only depends on the combination $m_\chi \cdot v_0$. Consequently, for those interaction types $N_\text{max}^{(\text{CDMS-Si})}$ approaches a constant in the limit $m_\chi \rightarrow 0$, as any further decrease of the dark matter mass can be compensated by a corresponding increase in the stream velocity.}.

As is well known, the CDMS-Si signal is incompatible with the null search results from XENON1T and PICO-60, if the scattering is mediated by the isoscalar, short-ranged, spin-independent interaction and the velocity distribution has the Maxwell-Boltzmann form;  for this case we obtain a maximum number of events of $\sim 3\times 10^{-3}$, which excludes a dark matter interpretation of the three CDMS-Si events by $\sim 6\sigma$. The number of events can be increased to $\sim 0.1$ if short-ranged isovector interactions are also included, and to $\sim 0.2$, if also long-range interactions are allowed. If no assumptions are done about the dark matter velocity distribution, our approach gives a maximum number of events of $\sim 0.03$, $0.2$ and $0.3$, respectively. The interpretation of the three CDMS-Si events in terms of the spin-independent interaction only is therefore excluded by at least $\sim 3\sigma$.

For arbitrary interactions of spin 0, spin 1/2 or spin 1 dark matter, which can be respectively mediated by 4, 14 or 16 non-relativistic operators, the dark matter interpretation of the three CDMS-Si events is marginally compatible with null results from XENON1T and PICO-60 for the Maxwell-Boltzmann distribution, when the interaction includes isoscalar and isovector, as well as short- and long-range interactions. More general velocity distributions relax the tension between the CDMS-Si data and the null results. Yet, the maximal number of events possible with the current constraints is just $\sim 5.4$, corresponding to a velocity distribution consisting of a single stream with velocity $\simeq 801\,$km/s~\footnote{As discussed above, in general the optimal solution can consist of two streams; numerically we find that in this and various other scenarios setting the weight of the second stream to zero leads to the optimal velocity distribution.}. Smearing the streams with a Gaussian distribution with variance $\lesssim$ 50 km/s does not change significantly the conclusions. For a larger smearing, the maximal number of events is again smaller than 3. An improvement in the sensitivity of both the xenon-based and fluorine-based experiments by a factor of $\simeq 20$ will suffice to exclude also these possibilities in a halo-independent manner at more than $3 \sigma$, or to the confirmation of the CDMS-Si signal with a different target material. In summary, even though for the concrete case of CDMS-Si a dark matter origin of the putative signal remains viable once one allows for arbitrary interactions and velocity distributions, our results clearly show that with the method presented in this work it is in principle possible to exclude a dark matter interpretation of an observed excess in a direct detection experiment fully independent of the velocity distribution, and nearly independent of the particle physics governing the dark matter-nucleon interaction.

Before concluding this section, we briefly comment on the relation between our work and recent studies relying on the expansion in Eq.~(\ref{eq:fv_form}), or, more in general, on Eq.~(\ref{eq:fdelta}).~As explained in Ref.~\cite{Gelmini:2017aqe}, the expansion of $f$ in a {\it finite} sum of delta functions, Eq.~(\ref{eq:fdelta}), is justified by Fenchel-Eggleston's theorem~\cite{Fenchel,Eggleston}, the extension of which to integral functionals, e.g.~Eq.~(\ref{eq:rate}), is based upon Choquet's theory~\cite{Choquet}.~Exploiting Eq.~(\ref{eq:fdelta}), linear programming algorithms were used to optimise physical observables which are linear functionals of $f$ given a set of constraints also linear in $f$~\cite{Ibarra:2017mzt}.~This approach can be used to derive a halo-independent upper limit on the dark matter-nucleon scattering cross-section from a set of null results; to confront a detection claim to a set of null results in a halo-independent manner; and to assess, also in a halo-independent manner, the prospects for dark matter detection in a future experiment given a set of current null results.~Furthermore, an expansion of $f$ in a finite sum of streams was used to extract the unmodulated nuclear recoil rate associated with the modulation signal reported by DAMA, with errors quantified through a profile likelihood approach~\cite{Gondolo:2017jro}.~This result was achieved through the optimisation of the unmodulated signal expected in a NaI detector subject to constraints from the modulation signal observed in DAMA.~Unlike these recent studies, in our work we used the expansion in Eq.~(\ref{eq:fdelta}) to optimise the number of signal events at CDMS-Si given the constraints from the null result of XENON1T and PICO-60.~The optimisation method used in our analysis is also new, and combines Eq.~(\ref{eq:zeta_condition}) and Eq.~(\ref{eq:fv_form}) with the analytic formula in appendix~\ref{app:vel_distr}.~As a result, within the non-relativistic effective theory of dark matter-nucleon interactions extended to massless mediators, we were able to perform the first halo-independent assessment of the compatibility of CDMS-Si with the null result of XENON1T and PICO-60.

\section{Conclusions}
\label{sec:conclusions}

We have presented a method to determine the compatibility of the dark matter interpretation of a tentative signal at a direct detection experiment with the null results from other experiments, addressing simultaneously the particle physics uncertainties on the concrete form of the dark matter-nucleon interaction, as well as the astrophysical uncertainties on the local dark matter density and velocity distribution. For the former, the method allows to include an arbitrarily large set of operators inducing scattering, which may and may not interfere with each other, and determines whether that set of operators provides, for a given velocity distribution, a common framework to reproduce all experimental results. For the latter, we determine analytically the velocity distribution which yields the largest rate compatible with the constraints from the null search experiments, which is a superposition of dark matter streams with fixed velocities. Thus, the problem of sampling over continuous velocity distributions reduces to a scan over a small number of parameters. 

We have applied the method to examine the dark matter interpretation of the three events observed in the silicon detectors of the CDMS-II experiment, in view of the null results from XENON1T and PICO-60, which is excluded assuming a isoscalar, short-ranged, spin-independent interaction, and assuming a Maxwell-Boltzmann velocity distribution. We find that the interpretation of the CDMS-Si events in terms of the spin-independent interaction is excluded, even when allowing for isovector interactions, long-range interactions, regardless of the velocity distribution. When allowing for more operators inducing the scattering, the limits from null search experiments get relaxed. Yet, we find that current constraints restrict the number of events at CDMS-Si to be smaller than $\sim 5.4$, for spin-1/2 or spin-1 dark matter, or  $\sim 4.2$ for spin-0 dark matter elastically scattering off nuclei. Modest improvements in sensitivity in experiments would lead either to a confirmation of the CDMS-Si signal, or to complete exclusion.  

\acknowledgments
This work has been supported by the DFG cluster of excellence EXC 153 ``Origin and Structure of the Universe'', by the Collaborative Research Center SFB1258, and by the ERC Starting Grant `NewAve' (638528). Finally, we would like to thank the participants of the programme ``Astro-, Particle and Nuclear Physics of Dark Matter Direct Detection'', hosted by MIAPP, for many valuable discussions.

\appendix

\section{Analysis of direct detection experiments}
\label{sec:exp}

In this appendix we describe the implementation of the results of the direct detection experiments employed in this work.\\
\textbf{XENON1T.} We calculate the total number of expected events in the XENON1T signal region using an exposure of $1042 \times 34.2 \, \text{kg}\cdot\text{days}$~\cite{Aprile:2017iyp}, and employing the acceptance function for nuclear recoils provided in the \texttt{DDCalc} package~\cite{Workgroup:2017lvb}.  The latter has been obtained by simulating fluctuations in the $S1$ and $S2$ signal using appropriate scintillation and ionization yields and then applying the corresponding analysis cuts as defined in~\cite{Aprile:2017iyp}; for details we refer to~\cite{Workgroup:2017lvb}. No events have been observed in the signal region of XENON1T, and by conservatively setting also the number of expected background events to zero, we obtain an upper limit of 3.00 signal events at $95\%\,$C.L. This procedure gives rise to an upper limit on the standard spin-independent scattering cross section which is a factor $\simeq 2 -3$ more conservative than the published result~\cite{Aprile:2017iyp}, rendering our results regarding the CDMS-Si excess somewhat conservative.\\ 
\textbf{PICO-60 (2017 data release).} We employ the most recent data release of the PICO experiment~\cite{Amole:2017dex} which uses a C$_3$F$_8$ target with an exposure of $1167\, \text{kg}\cdot\text{days}$. The acceptance as a function of the nuclear recoil energy is obtained from the black and red dashed curves in Fig.~4 of~\cite{Amole:2015lsj}, shifted by 0.1 keV towards larger recoil energies in order to take into account the slightly increased threshold of the analysis in~\cite{Amole:2017dex} compared to~\cite{Amole:2015lsj}. As for XENON1T no events have been observed, leading to the $95\%\,$C.L. upper limit of 3.00 expected events. Our corresponding upper limit on the spin-dependent scattering cross section of dark matter with protons is in agreement at the percent level with the published result at $m_\chi \gtrsim 10\,$GeV, while it is slightly more constraining at smaller masses.\\
\textbf{CDMS-Si}
Our implementation of CDMS-Si is based on an exposure of $140.2\,\text{kg}\cdot\text{days}$, using as efficiency the blue solid curve shown in Fig.~1 of~\cite{Agnese:2013rvf}. Three events have been observed at reconstructed recoil energies of 8.2 keV, 9.5 keV and 12.3 keV. As we are interested in the dark matter induced events potentially explaining those, we only integrate the differential recoil rate over the energy range $[7.6\,\text{keV},\,12.9\,\text{keV}]$. With this choice we take into account energy fluctuations of up to $2\sigma_E$, assuming $\sigma_E = 0.3\,$keV~\cite{Akerib:2010pv,Frandsen:2013cna}.

\section{Optimized velocity distributions for CDMS-Si}
\label{app:vel_distr}
We decompose the velocity distribution in the detector rest frame as a linear superposition of $n$ streams with fixed velocity:
\begin{align}
f(\vec v)=\sum_{i=1}^n a_{\vec v_i}^2 \delta(\vec v- \vec v_i)
\label{eq:fdelta}
\end{align}
where $n$ can be arbitrarily large, and $a_{\vec v_i}^2$ is the (non-negative) weight of the stream with velocity $\vec v_i$. With this decomposition, the number of events at the direct detection experiment ${\cal E}$ can be cast as:
\begin{align}
N_{f(\vec v)}^{({\cal E})} \left( {\bf c }\right) 
= \sum_{i=1}^n a_{\vec v_i}^2 N^{({\cal E})}_{\vec v_i} ({\bf c})
=\sum_{i=1}^n \sum_{p,q=1}^D a_{\vec v_i}^2 c_p \left(\mathbb{N}_{\vec v_i}^{({\cal E})}\right)_{pq} c_q
\end{align}
with $N^{({\cal E})}_{\vec v_i} ({\bf c})$ the number of expected signal events for the set of coupling constants ${\bf c}=\{ c_p\},$ $p=1,...,D$, should the velocity distribution be a stream with fixed velocity $\vec v_i$.

The optimization problem can then be formulated as:
\begin{align}
{\rm maximize}&~~ \sum_{i=1}^n \sum_{p,q=1}^D a_{\vec v_i}^2 c_p \left(\varmathbb{N}^{\rm (CDMS-Si)}_{\vec v_i}\right)_{pq} c_q\;, \nonumber \\
{\rm subject~to}&~~\sum_{i=1}^n \sum_{p,q=1}^D a_{\vec v_i}^2 c_p \left(\varmathbb{N}^{\rm (XENON1T)}_{\vec v_i}\right)_{pq} c_q\leq N^{\rm (XENON1T)}_{\rm u.l.}\;, \nonumber \\
{\rm and}&~~ \sum_{i=1}^n\sum_{p,q=1}^D a_{\vec v_i}^2 c_p \left(\varmathbb{N}^{\rm (PICO)}_{\vec v_i}\right)_{pq} c_q\leq N^{\rm (PICO)}_{\rm u.l.}\;, \nonumber \\
{\rm and}&~~\sum_{i=1}^n a^2_{\vec v_i}=1\;.
\label{eq:maximization_problem}
\end{align}

In order to minimize the objective function with constraints it is convenient to introduce the Lagrangian
\begin{align}
L(\{a_{\vec v_i}\}, \{\vec v_i\}, \{c_p\}, s_1, s_2, \lambda_1,\lambda_2,\lambda_3) &=\sum_{i=1}^n \sum_{p,q=1}^D a_{\vec v_i}^2 c_p (\varmathbb{N}^{\rm (CDMS-Si)}_{\vec v_i})_{pq} c_q \nonumber \\
&-\lambda_1\Big( \sum_{i=1}^n \sum_{p,q=1}^D a_{\vec v_i}^2 c_p (\varmathbb{N}^{\rm (XENON1T)}_{\vec v_i})_{pq} c_q+s_1^2- N^{\rm XENON1T}_{\rm u.l.} \Big)\nonumber \\
&-\lambda_2\Big( \sum_{i=1}^n \sum_{p,q=1}^D a_{\vec v_i}^2 c_p (\varmathbb{N}^{\rm (PICO)}_{\vec v_i})_{pq} c_q+s_2^2- N^{\rm PICO}_{\rm u.l.}  \Big)\nonumber  \\
&-\lambda_3\Big( \sum_{i=1}^n  a_{\vec v_i}^2 -1 \Big) \,,
\end{align}
where $\{\vec v_i\}\equiv \{\vec v_1,...\vec v_n\}$ and $\{a_{\vec v_i}\}\equiv \{a_{\vec v_1},..., a_{\vec v_n}\}$
denote the velocities and weights of the $n$ streams used in the decomposition of the velocity distribution, $\{c_p\}\equiv \{c_1,..., c_D\}$ denotes the coefficients of the $D$ effective operators used in our effective theory approach,  $\lambda_1$, $\lambda_2$ and $\lambda_3$ are Lagrange multipliers, and $s_1^2$ and  $s_2^2$ are (non-negative) slack variables, introduced  to recast the upper inequality constraints into equality constraints. The maximization conditions are:
\begin{align}
\frac{\partial L}{\partial a_{\vec v_j}}&=2 a_{\vec v_j}\left\{ \sum_{p,q=1}^D  c_p \Big[ \varmathbb{N}^{\rm (CDMS-Si)}_{\vec v_j} -\lambda_1 \varmathbb{N}^{\rm (XENON1T)}_{\vec v_j}-\lambda_2 \varmathbb{N}^{\rm (PICO)}_{\vec v_j}\Big]_{pq} c_q -\lambda_3 \right\}=0
,~~~ j=1, ..., n \;,
\label{eq:cond1}\\
\frac{\partial L}{\partial \vec v_j}&= a_{\vec v_j}^2 \sum_{p,q=1}^D  c_p \left[ \left(\frac{\partial \varmathbb{N}^{\rm (CDMS-Si)}_{\vec v_j}}{\partial \vec v_j}\right) -\lambda_1 \left(\frac{\partial \varmathbb{N}^{\rm (XENON1T)}_{\vec v_j}}{\partial \vec v_j}\right)-\lambda_2\left(\frac{\partial \varmathbb{N}^{\rm (PICO)}_{\vec v_j}}{\partial \vec v_j}\right)\right]_{pq} c_q=0,~~~ j=1, ..., n \;,
\label{eq:cond2}\\
\frac{\partial L}{\partial s_1}&=2 \lambda_1 s_1 =0\;, 
\label{eq:cond3}\\
\frac{\partial L}{\partial s_2}&=2 \lambda_2 s_2 =0\;, 
\label{eq:cond4}\\
\frac{\partial L}{\partial \lambda_1}&= \sum_{i=1}^n \sum_{p,q=1}^D a_{\vec v_i}^2 c_p (\varmathbb{N}^{\rm (XENON1T)}_{\vec v_i})_{pq} c_q+s_1^2- N^{\rm (XENON1T)}_{\rm u.l.}=0\;, 
\label{eq:cond5}\\
\frac{\partial L}{\partial \lambda_2}&= \sum_{i=1}^n \sum_{p,q=1}^D a_{\vec v_i}^2 c_p (\varmathbb{N}^{\rm (PICO)}_{\vec v_i})_{pq} c_q+s_2^2- N^{\rm (PICO)}_{\rm u.l.}=0\;, 
\label{eq:cond6}\\
\frac{\partial L}{\partial \lambda_3}&=\sum_{i=1}^n a_{\vec v_i}^2  -1=0\;, 
\label{eq:cond7}\\
\frac{\partial L}{\partial c_p}&= \sum_{i=1}^n \sum_{q=1}^D 2 a^2_{\vec v_i}   \Big[ \varmathbb{N}^{\rm (CDMS-Si)}_{\vec v_i} -\lambda_1 \varmathbb{N}^{\rm (XENON1T)}_{\vec v_i}-\lambda_2 \varmathbb{N}^{\rm (PICO)}_{\vec v_i}\Big]_{pq} c_q=0,~~~ p=1..., D\;.
\label{eq:cond8}
\end{align}
We first note that for the parameters that maximize the Lagrangian one has
\begin{align}
\sum_{j=1}^n a_{\vec v_j}\frac{\partial L}{\partial a_{\vec v_j}}-\sum_{p=1}^D c_p \frac{\partial L}{\partial c_p}=0 \,.
\end{align}
Substituting Eq.~(\ref{eq:cond1}) and  Eq.~(\ref{eq:cond8}) as well as using Eq.~(\ref{eq:cond7}) we obtain
\begin{align}
0=-2\lambda_3 \left(\sum_{i=1}^n a_{\vec v_i}^2 \right)=-2\lambda_3\;,
\end{align}
from where it follows that $\lambda_3= 0$.

Besides, Eq.~(\ref{eq:cond3})  is satisfied either when $s_1=0$ or when $\lambda_1=0$, and Eq.~(\ref{eq:cond4})  is satisfied either when $s_2=0$ or when $\lambda_2=0$. A vanishing slack variable implies that the upper limit on the number of signal events is saturated. Therefore, $\lambda_1\neq 0$ ($\lambda_2\neq 0$) implies that, for the parameters that maximize the number of signal events at CDMS-Si, the upper bounds from XENON1T (PICO-60) are saturated.  while  $\lambda_1\neq 0$ and $\lambda_2\neq 0$ imply that the upper bounds from XENON1T and PICO-60 are simultaneously saturated.

In the case where both $\lambda_1$ and $\lambda_2$ are non-vanishing,  Eq.~(\ref{eq:cond1}) reads
\begin{align}
a_{\vec v_j}\left(\langle \varmathbb{N}^{\rm (CDMS-Si)}_{\vec v_j}\rangle -\lambda_1 \langle \varmathbb{N}^{\rm (XENON1T)}_{\vec v_j}\rangle -\lambda_2 \langle \varmathbb{N}^{\rm (PICO)}_{\vec v_j}\rangle\right)=0,~~~ j=1...,n \,,
\label{eq:c_as_eigenvector}
\end{align}
with $\langle ...\rangle \equiv\langle {\bf c}|...|{\bf c}\rangle$. Since we only have two non-vanishing Lagrange multipliers, it follows that these equations can only be satisfied if 
\begin{align}
&\langle \varmathbb{N}^{\rm (CDMS-Si)}_{\vec v_1}\rangle -\lambda_1 \langle \varmathbb{N}^{\rm (XENON1T)}_{\vec v_1}\rangle -\lambda_2 \langle \varmathbb{N}^{\rm (PICO)}_{\vec v_1}\rangle=0 \,,\\
&\langle \varmathbb{N}^{\rm (CDMS-Si)}_{\vec v_2}\rangle -\lambda_1 \langle \varmathbb{N}^{\rm (XENON1T)}_{\vec v_2}\rangle -\lambda_2 \langle \varmathbb{N}^{\rm (PICO)}_{\vec v_2}\rangle=0 \,, \\
&a_{\vec v_j}=0\,,~~~~~~ j=3,...n \,,
\end{align}
for some $\vec v_1$, $\vec v_2$\footnote{In case not all of the equations~(\ref{eq:c_as_eigenvector}) are linearly independent, there are also solutions with more than two non-vanishing $a_{\vec v_j}$. However, it is straightforward to show that in such a scenario there is always a configuration with only two non-zero $a_{\vec v_j}$ giving rise to the same number of expected events in all three experiments.}. Therefore, the optimized velocity distribution corresponds to a superposition of two streams with weights $a_{\vec v_1}$ and $a_{\vec v_2}$ satisfying $a_{\vec v_1}^2+a_{\vec v_2}^2=1$, which can be cast as
\begin{align}
f(\vec v)=\alpha \delta(\vec v -\vec v_1)+(1-\alpha) \delta(\vec v -\vec v_2)
\label{eq:optimized_distribution}
\end{align}
which $0 \leq\alpha \leq 1$. If one of the Lagrange multipliers vanishes (which corresponds to the case where the upper limit on the number of events at XENON1T or PICO-60 is not saturated), then an analogous calculation shows that the optimized velocity distribution corresponds to just one stream. This case is still described by Eq.~(\ref{eq:optimized_distribution}) with $\alpha=0$ or 1. 

To summarize, the velocity distribution that solves the optimization problem Eq.~(\ref{eq:maximization_problem}) has the form Eq.~(\ref{eq:optimized_distribution}) with $0\leq \alpha \leq 1$.

\bibliographystyle{JHEP-mod}
\bibliography{refs}

\providecommand{\href}[2]{#2}\begingroup\raggedright\begin{thebibliography}{10}

\bibitem{Bertone:2004pz}
G.~Bertone, D.~Hooper, and J.~Silk, {\it {Particle dark matter: Evidence,
  candidates and constraints}},  {\em Phys. Rept.} {\bf 405} (2005) 279--390,
  [\href{http://xxx.lanl.gov/abs/hep-ph/0404175}{{\tt hep-ph/0404175}}].

\bibitem{Goodman:1984dc}
M.~W. Goodman and E.~Witten, {\it {Detectability of Certain Dark Matter
  Candidates}},  {\em Phys. Rev.} {\bf D31} (1985) 3059.

\bibitem{Aprile:2017iyp}
{\bf XENON}, E.~Aprile {\em et.~al.}, {\it {First Dark Matter Search Results
  from the XENON1T Experiment}},  {\em Phys. Rev. Lett.} {\bf 119} (2017),
  no.~18 181301, [\href{http://xxx.lanl.gov/abs/1705.06655}{{\tt
  arXiv:1705.06655}}].

\bibitem{Cui:2017nnn}
{\bf PandaX-II}, X.~Cui {\em et.~al.}, {\it {Dark Matter Results From
  54-Ton-Day Exposure of PandaX-II Experiment}},  {\em Phys. Rev. Lett.} {\bf
  119} (2017), no.~18 181302, [\href{http://xxx.lanl.gov/abs/1708.06917}{{\tt
  arXiv:1708.06917}}].

\bibitem{Akerib:2016vxi}
{\bf LUX}, D.~S. Akerib {\em et.~al.}, {\it {Results from a search for dark
  matter in the complete LUX exposure}},  {\em Phys. Rev. Lett.} {\bf 118}
  (2017), no.~2 021303, [\href{http://xxx.lanl.gov/abs/1608.07648}{{\tt
  arXiv:1608.07648}}].

\bibitem{Amole:2017dex}
{\bf PICO}, C.~Amole {\em et.~al.}, {\it {Dark Matter Search Results from the
  PICO-60 C$_3$F$_8$ Bubble Chamber}},  {\em Phys. Rev. Lett.} {\bf 118}
  (2017), no.~25 251301, [\href{http://xxx.lanl.gov/abs/1702.07666}{{\tt
  arXiv:1702.07666}}].

\bibitem{Ahmed:2010wy}
{\bf CDMS-II}, Z.~Ahmed {\em et.~al.}, {\it {Results from a Low-Energy Analysis
  of the CDMS II Germanium Data}},  {\em Phys. Rev. Lett.} {\bf 106} (2011)
  131302, [\href{http://xxx.lanl.gov/abs/1011.2482}{{\tt arXiv:1011.2482}}].

\bibitem{Agnese:2013rvf}
{\bf CDMS}, R.~Agnese {\em et.~al.}, {\it {Silicon Detector Dark Matter Results
  from the Final Exposure of CDMS II}},  {\em Phys. Rev. Lett.} {\bf 111}
  (2013), no.~25 251301, [\href{http://xxx.lanl.gov/abs/1304.4279}{{\tt
  arXiv:1304.4279}}].

\bibitem{Angloher:2015ewa}
{\bf CRESST}, G.~Angloher {\em et.~al.}, {\it {Results on light dark matter
  particles with a low-threshold CRESST-II detector}},  {\em Eur. Phys. J.}
  {\bf C76} (2016), no.~1 25, [\href{http://xxx.lanl.gov/abs/1509.01515}{{\tt
  arXiv:1509.01515}}].

\bibitem{Frandsen:2013cna}
M.~T. Frandsen, F.~Kahlhoefer, C.~McCabe, S.~Sarkar, and K.~Schmidt-Hoberg,
  {\it {The unbearable lightness of being: CDMS versus XENON}},  {\em JCAP}
  {\bf 1307} (2013) 023, [\href{http://xxx.lanl.gov/abs/1304.6066}{{\tt
  arXiv:1304.6066}}].

\bibitem{DelNobile:2013cva}
E.~Del~Nobile, G.~Gelmini, P.~Gondolo, and J.-H. Huh, {\it {Generalized Halo
  Independent Comparison of Direct Dark Matter Detection Data}},  {\em JCAP}
  {\bf 1310} (2013) 048, [\href{http://xxx.lanl.gov/abs/1306.5273}{{\tt
  arXiv:1306.5273}}].

\bibitem{Fox:2014kua}
P.~J. Fox, Y.~Kahn, and M.~McCullough, {\it {Taking Halo-Independent Dark
  Matter Methods Out of the Bin}},  {\em JCAP} {\bf 1410} (2014), no.~10 076,
  [\href{http://xxx.lanl.gov/abs/1403.6830}{{\tt arXiv:1403.6830}}].

\bibitem{Scopel:2014kba}
S.~Scopel and K.~Yoon, {\it {A systematic halo-independent analysis of direct
  detection data within the framework of Inelastic Dark Matter}},  {\em JCAP}
  {\bf 1408} (2014) 060, [\href{http://xxx.lanl.gov/abs/1405.0364}{{\tt
  arXiv:1405.0364}}].

\bibitem{Feldstein:2014ufa}
B.~Feldstein and F.~Kahlhoefer, {\it {Quantifying (dis)agreement between direct
  detection experiments in a halo-independent way}},  {\em JCAP} {\bf 1412}
  (2014), no.~12 052, [\href{http://xxx.lanl.gov/abs/1409.5446}{{\tt
  arXiv:1409.5446}}].

\bibitem{Bozorgnia:2014gsa}
N.~Bozorgnia and T.~Schwetz, {\it {What is the probability that direct
  detection experiments have observed Dark Matter?}},  {\em JCAP} {\bf 1412}
  (2014), no.~12 015, [\href{http://xxx.lanl.gov/abs/1410.6160}{{\tt
  arXiv:1410.6160}}].

\bibitem{Anderson:2015xaa}
A.~J. Anderson, P.~J. Fox, Y.~Kahn, and M.~McCullough, {\it {Halo-Independent
  Direct Detection Analyses Without Mass Assumptions}},  {\em JCAP} {\bf 1510}
  (2015), no.~10 012, [\href{http://xxx.lanl.gov/abs/1504.03333}{{\tt
  arXiv:1504.03333}}].

\bibitem{Gelmini:2016pei}
G.~B. Gelmini, J.-H. Huh, and S.~J. Witte, {\it {Assessing Compatibility of
  Direct Detection Data: Halo-Independent Global Likelihood Analyses}},  {\em
  JCAP} {\bf 1610} (2016), no.~10 029,
  [\href{http://xxx.lanl.gov/abs/1607.02445}{{\tt arXiv:1607.02445}}].

\bibitem{Gondolo:2017jro}
P.~Gondolo and S.~Scopel, {\it {Halo-independent determination of the
  unmodulated WIMP signal in DAMA: the isotropic case}},
  \href{http://xxx.lanl.gov/abs/1703.08942}{{\tt arXiv:1703.08942}}.

\bibitem{Ibarra:2017mzt}
A.~Ibarra and A.~Rappelt, {\it {Optimized velocity distributions for direct
  dark matter detection}},  \href{http://xxx.lanl.gov/abs/1703.09168}{{\tt
  arXiv:1703.09168}}.

\bibitem{Catena:2016hoj}
R.~Catena, A.~Ibarra, and S.~Wild, {\it {DAMA confronts null searches in the
  effective theory of dark matter-nucleon interactions}},  {\em JCAP} {\bf
  1605} (2016), no.~05 039, [\href{http://xxx.lanl.gov/abs/1602.04074}{{\tt
  arXiv:1602.04074}}].

\bibitem{Fan:2010gt}
J.~Fan, M.~Reece, and L.-T. Wang, {\it {Non-relativistic effective theory of
  dark matter direct detection}},  {\em JCAP} {\bf 1011} (2010) 042,
  [\href{http://xxx.lanl.gov/abs/1008.1591}{{\tt arXiv:1008.1591}}].

\bibitem{Fitzpatrick:2012ix}
A.~L. Fitzpatrick, W.~Haxton, E.~Katz, N.~Lubbers, and Y.~Xu, {\it {The
  Effective Field Theory of Dark Matter Direct Detection}},  {\em JCAP} {\bf
  1302} (2013) 004, [\href{http://xxx.lanl.gov/abs/1203.3542}{{\tt
  arXiv:1203.3542}}].

\bibitem{Dent:2015zpa}
J.~B. Dent, L.~M. Krauss, J.~L. Newstead, and S.~Sabharwal, {\it {General
  analysis of direct dark matter detection: From microphysics to observational
  signatures}},  {\em Phys. Rev.} {\bf D92} (2015), no.~6 063515,
  [\href{http://xxx.lanl.gov/abs/1505.03117}{{\tt arXiv:1505.03117}}].

\bibitem{Anand:2013yka}
N.~Anand, A.~L. Fitzpatrick, and W.~Haxton, {\it {Model-independent WIMP
  Scattering Responses and Event Rates: A Mathematica Package for Experimental
  Analysis}},  {\em Phys. Rev.} {\bf C89} (2014) 065501,
  [\href{http://xxx.lanl.gov/abs/1308.6288}{{\tt arXiv:1308.6288}}].

\bibitem{Catena:2015uha}
R.~Catena and B.~Schwabe, {\it {Form factors for dark matter capture by the Sun
  in effective theories}},  {\em JCAP} {\bf 1504} (2015), no.~04 042,
  [\href{http://xxx.lanl.gov/abs/1501.03729}{{\tt arXiv:1501.03729}}].

\bibitem{Warburton:1992rh}
E.~Warburton and B.~A. Brown, {\it {Effective Interactions for the Op1sOd
  nuclear shell model space}},  {\em Phys. Rev.} {\bf C46} (1992) 923--944.

\bibitem{MATLAB:2014a}
MATLAB, {\em version 8.3.0.532 (R2014a)}.
\newblock The MathWorks Inc., Natick, Massachusetts, 2014.

\bibitem{Gelmini:2017aqe}
G.~B. Gelmini, J.-H. Huh, and S.~J. Witte, {\it {Unified Halo-Independent
  Formalism Derived From Convex Hulls}},
  \href{http://xxx.lanl.gov/abs/1707.07019}{{\tt arXiv:1707.07019}}.

\bibitem{Fenchel}
W.~Fenchel, {\em Convex cones, sets and functions}.
\newblock Department of Mathematics, Princeton University, 1953.

\bibitem{Eggleston}
H.~Eggleston, {\em Convexity (Cambridge Tracts in Mathematics and Mathematical
  Physics. No.47)}.
\newblock Cambridge University Press, 1958.

\bibitem{Choquet}
G.~Choquet, {\it Existence et unicit{\'e} des repr{\'e}sentations
  int{\'e}grales au moyen des points extr{\'e}maux dans les c{\^o}nes
  convexes},  vol.~4, pp.~33--47, S{\'e}minaire Bourbaki, 1956.

\bibitem{Workgroup:2017lvb}
{\bf GAMBIT Dark Matter Workgroup}, T.~Bringmann {\em et.~al.}, {\it {DarkBit:
  A GAMBIT module for computing dark matter observables and likelihoods}},
  \href{http://xxx.lanl.gov/abs/1705.07920}{{\tt arXiv:1705.07920}}.

\bibitem{Amole:2015lsj}
{\bf PICO}, C.~Amole {\em et.~al.}, {\it {Dark Matter Search Results from the
  PICO-2L C$_3$F$_8$ Bubble Chamber}},  {\em Phys. Rev. Lett.} {\bf 114}
  (2015), no.~23 231302, [\href{http://xxx.lanl.gov/abs/1503.00008}{{\tt
  arXiv:1503.00008}}].

\bibitem{Akerib:2010pv}
{\bf CDMS}, D.~S. Akerib {\em et.~al.}, {\it {A low-threshold analysis of CDMS
  shallow-site data}},  {\em Phys. Rev.} {\bf D82} (2010) 122004,
  [\href{http://xxx.lanl.gov/abs/1010.4290}{{\tt arXiv:1010.4290}}].

\end{thebibliography}\endgroup

\end{document}